\documentstyle[times,graphicx,amssymb,amsmath,prb,aps]{revtex}

\begin{document}

\draft % Revtex3 -- prints PACS numbers

\wideabs{% Revtex3 -- two column layout

  \title{The effect of electromechanical coupling on the strain in
    AlGaN/GaN heterojunction field effect transistors}

\author{B. Jogai}
% \email[e-mail: ]{brahmanand.jogai@wpafb.af.mil} % Revtex4
% \affiliation{Air Force Research Laboratory, Materials
%   and Manufacturing Directorate, Wright-Patterson Air Force Base, OH
%   45433}
% \affiliation{Semiconductor Research Center, Wright State University,
%   Dayton, OH 45435}
\address{Air Force Research Laboratory, Materials
   and Manufacturing Directorate, Wright-Patterson Air Force Base,
   Ohio
   45433 and Semiconductor Research Center, Wright State University,
   Dayton, Ohio 45435}
\author{J. D. Albrecht}
\address{Air Force Research Laboratory, Wright-Patterson Air Force
   Base, Ohio 45433}
\author{E. Pan}
\address{Department of Civil Engineering, The University of Akron,
   Akron, Ohio 44325}

\maketitle % Revtex3

\begin{abstract}
  The strain in AlGaN/GaN heterojunction field-effect transistors
  (HFETs) is examined theoretically in the context of the
  fully-coupled equation of state for piezoelectric materials.  Using
  a simple analytical model, it is shown that, in the absence of a
  two-dimensional electron gas (2DEG), the out-of-plane strain
  obtained without electromechanical coupling is in error by about
  30\% for an Al fraction of 0.3.  This result has consequences for
  the calculation of quantities that depend directly on the strain
  tensor.  These quantities include the eigenstates and electrostatic
  potential in AlGaN/GaN heterostructures.  It is shown that for an
  HFET, the electromechanical coupling is screened by the 2DEG.
  Results for the electromechanical model, including the 2DEG,
  indicate that the standard (decoupled) strain model is a reasonable
  approximation for HFET calculataions.  The analytical results are
  supported by a self-consistent Schr\"odinger-Poisson calculation
  that includes the fully-coupled equation of state together with the
  charge-balance equation.
\end{abstract}

}

\section{Introduction}
\label{sec:intro} The piezoelectric and spontaneous polarization
properties of the Al$_x$Ga$_{1-x}$N material system makes it
attractive for certain wide bandgap device applications.
Heterojunction field-effect transistors (HFETs) made from
AlGaN/GaN possess two-dimensional electron gas (2DEG)
concentrations on the order of $10^{13}$ $ \textrm{cm}^{-2}$ at
the AlGaN/GaN interface. Owing to the polarization effects, the
2DEG can be established at zero gate bias and achieved without
intentional
doping.\cite{Ambacher1,Antoszewski,Zhang,Ibbetson,Rumyantsev2,Jiang}
A back doping design\cite{Maeda} can further increase the 2DEG
concentration to approximately $ 3\times 10^{13} $ $
\textrm{cm}^{-2} $. The dense 2DEGs result in r.f. device power
densities in the range 6 -- 12 W/mm, making AlGaN/GaN HFETs
attractive for high-power applications.

The formation of the 2DEG can be traced directly to the large
built-in polarization fields in the material system.  In addition
to the piezoelectric polarization due to the strain induced by the
lattice mismatch, there is a spontaneous polarization in the
direction $ \textrm{N} \rightarrow \textrm{Ga} $ along the
$c$-axis which can be several times larger than the piezoelectric
polarization.  These polarizations induce local electric fields
which, in turn, cause a notch to form in the conduction band edge
at the interface.  Electrons congregate in the notch, either from
surface states or from unintentional dopants and traps, forming
the 2DEG.

In previous HFET analyses, the electrical and mechanical
properties of the epitaxial layers have been treated
independently. The standard elastic theory is applied and Hooke's
law is assumed to hold in obtaining the strain tensors in each
material layer. In effect, this assumption decouples the strain
tensor from the Poisson and Schr\"odinger equations and renders
the task of computing the free charge distribution and band
structure more tractable.  In reality, however, the electrical and
mechanical properties of piezoelectric materials are coupled as
seen from thermodynamics\cite{Nye} and are best treated using a
unified approach, especially since the piezoelectric response is
large in AlGaN for high Al fractions.

In other work, large modifications to the elastic properties have
been predicted for buried group III-nitride
nanostructures\cite{Pan2} and free-standing plates of
AlN.\cite{Pan1} The latter is a geometry closely related to the
AlGaN barrier layer of the HFET. As a direct comparison, we take
an AlN plate experiencing biaxial strain equivalent to the strain
of a pseudomorphic AlN layer on GaN, or
$\gamma_{xx}$$=$$\gamma_{yy}$$=$$-0.024$. The resulting strain
using an uncoupled calculation in the absence of spontaneous
polarization is $\gamma_{zz}$$=$$0.014$ and the electric field is
$E_z$$=$$-5.1$ MV/cm. In the coupled calculation, we obtain
$\gamma_{zz}$$=$$0.012$ and $E_z$$=$$-4.7$ MV/cm. Based on these
substantial differences there is strong motivation to make a
detailed analysis of a transistor structure including a detailed
analysis of the spontaneous polarization moment and free carrier
distributions.

In the present work, the strain and electric fields obtained from
the fully-coupled model are contrasted with those from the
standard model for a representative HFET structure. The deviations
from the standard model are discussed in detail. The theory uses
the fully-coupled piezoelectric equation of state to obtain the
strain field. A simple analytical model is then used to obtain the
interaction between the polarization-induced electric field and
the strain tensor.  Also included is the effect of 2DEG
charge-screening on the electromechanical coupling.  The results
from the analytical model are supported by those from a
self-consistent Schr\"odinger-Poisson model that includes charge
balance and free hole states.

This paper is organized as follows: In Sec.\ \ref{sec:method}, the
strain tensor is derived from the fully-coupled equation of state.
The Poisson equation is solved analytically for a bi-layer
AlGaN/GaN slab and the strain components and electric field in the
barrier material is derived in closed form.  In Sec.\
\ref{sec:results}, the calculated deviations of the standard model
relative to the fully-coupled model are shown. The effect of
screening from the 2DEG is discussed. The results from a
fully-coupled numerical model are presented.  The results are
summarized in Sec.\ \ref{sec:summary}.

\section{Model description}
\label{sec:method}
\begin{figure}
  \begin{center}
    \includegraphics[width=3.2in]{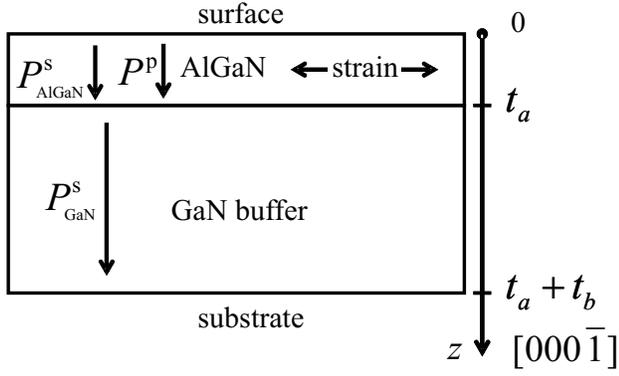}
  \end{center}
  \caption{Cross-section of a model cation-faced AlGaN/GaN HFET
    showing the direction of the piezoelectric and spontaneous
    polarization vectors in relation to the $z$-axis. $t_a$ and $t_b$
    are the thicknesses of the AlGaN and GaN layers, respectively.}
  \label{fig:schematic}
\end{figure}

The fully-coupled equation of state for piezoelectric materials is
expressed as\cite{Zelenka}
\begin{equation}
  \label{eq:state}
  \sigma_{ij} = C_{ijkl} \gamma_{kl} - e_{kij} E_k ,
\end{equation}
in component form where $\sigma_{ij}$ is the stress tensor,
$C_{ijkl}$ is the fourth-ranked elastic stiffness tensor,
$\gamma_{kl}$ is the strain tensor, $e_{kij}$ is the third-ranked
piezoelectric coefficient tensor, $E_k$ is the electric field
vector, and the indices $i$, $j$, $k$, and $l$ run over the
Cartesian coordinates $x$, $y$, and $z$. Summation over repeated
indices has been suppressed. Eq.\ (\ref{eq:state}) in the absence
of piezoelectric coupling ($e_{kij}$$\rightarrow$$0$) is
recognized as Hooke's law relating stress to strain. The
fully-coupled model is now applied to the representative HFET
structure shown in Fig.\ \ref{fig:schematic}. A one-dimensional
(1-D) approach is used where the strain in the bi-layer slab is
assumed to be piecewise homogeneous.  This approximation means
that $\gamma_{ij}$ is diagonal for the coordinates defined in Fig.\
\ref{fig:schematic} and that $ \gamma_{xx} $ and $ \gamma_{yy} $
are, to first order, uniform in each layer. The strain induces a
piezoelectric polarization given by
\begin{equation}
  \label{eq:piezo}
  P^{\textrm{p}}_i = e_{ijk} \gamma_{jk} .
\end{equation}
From straightforward electrostatics we know that Eqs.\
(\ref{eq:state}) and (\ref{eq:piezo}) are coupled through the
Poisson equation which is written in our notation as
\begin{equation}
  \label{eq:poisdef}
  \nabla \cdot \epsilon  \nabla \phi  =  - \rho
   + \nabla \cdot (\textbf{P}^\textrm{s} +
  \textbf{P}^\textrm{p} ) ,
\end{equation}
where $ \phi $ is the electrostatic potential, $\epsilon$ is the
electric permittivity, $\rho$ is the charge distribution and
includes both ionic and free charges, $ \textbf{P}^\textrm{s} $ is
the spontaneous polarization, and $ \textbf{P}^\textrm{p} $ has
been defined in Eq.\ (\ref{eq:piezo}).  In the absence of shear
strains, $ \textbf{P}^\textrm{s} $ and $ \textbf{P}^\textrm{p} $
are directed in the $ [000\bar{1}] $ direction for a cation-faced
structure as shown in Fig.\ \ref{fig:schematic}, and with the
sixfold rotational symmetry along the $c$-axis, $\gamma_{xx}$$=$$
\gamma_{yy}$.  From Eq.\ (\ref{eq:piezo}), the piezoelectric
polarization along $[000\bar{1}]$ for wurtzite materials then
becomes
\begin{equation}
  \label{eq:piezo_final}
  P^\textrm{p} = 2 e_{31} \gamma_{xx} + e_{33} \gamma_{zz} .
\end{equation}
It is assumed that $ \gamma_{xx} $ is determined by the atomic
alignment in the $c$-plane across the AlGaN/GaN interface.  This
pseudomorphic boundary condition gives
\begin{equation}
  \label{eq:exx}
  \gamma_{xx} = \frac{a_b - a_a}{a_a} ,
\end{equation}
in the AlGaN layer and $\gamma_{xx}$$=$$0$ in the GaN layer, where
$ a_a $ and $ a_b $ are the $c$-plane lattice constants in the
AlGaN and GaN layers, respectively.  In reality, besides the force
exerted on the AlGaN layer by the GaN layer, one would also expect
the AlGaN layer to exert a force on the GaN layer, causing some
strain on the GaN side of the interface.  This local distortion of
the lattice would be a maximum at the junction and become
vanishingly small towards the interior of the GaN.  Such
refinements, however, would require a three-dimensional (3-D)
modeling of the electromechanical problem, as it would have to
include non-zero and position-dependent shear terms in the strain
tensor.  This is beyond the scope of the present work and is not
essential for our investigation.

Thus, within the present work, $ \gamma_{xx} $ is assumed to be
decoupled from the electrical properties, essentially fixed by the
bonding arrangement at the interfacial plane, with $ \gamma_{zz} $
serving as the principal vehicle for electromechanical coupling.
From Eq.\ (\ref{eq:state}) and from the foregoing assumptions, the
normal component of the stress tensor is given by
\begin{equation}
  \label{eq:sigmazz}
  \sigma_{zz} = 2 C_{13} \gamma_{xx} + C_{33} \gamma_{zz} - e_{33} E_z .
\end{equation}
From the boundary conditions for a free surface, the stresses along
the outward normal must be zero.  As a consequence, Eq.\
(\ref{eq:sigmazz}) then gives for the 1-D case
\begin{equation}
  \label{eq:ezz}
  \gamma_{zz} = -\frac{2 C_{13} }{C_{33}} \gamma_{xx} + \frac{e_{33}}{C_{33}}
  E_z
\end{equation}
which is the out-of-plane strain, where $E_z$$=$$- \partial \phi /
\partial z$. From Eq.\ (\ref{eq:ezz}), it is seen that the first term
is consistent with the Poisson ratio of the standard model for $
\gamma_{zz} $ and a new coupled term representing the
electromechanical coupling.  To date, the latter has not been included
in AlGaN/GaN HFET models.  The extent of the coupling will depend on
the Al mole fraction and the electric field in the layers, which, in
turn, will depend on the charge distribution and boundary conditions
used to solve Eq.\ (\ref{eq:poisdef}).  The actual problem is quite
complicated and involves the self-consistent solution of the equation
of state in conjunction with the Poisson, Schr\"odinger, and charge
balance equations.  To illustrate the concept of electromechanical
coupling without obscuring much of the underlying physics, a simple
example has been contrived from which analytical results can be
extracted so that the effect of coupling can be easily quantified.
The charges assumed herein are the polarization space charges and a
sheet electron distribution at the AlGaN/GaN interface.  Ordinarily,
the electron distribution should be calculated quantum-mechanically
from the electron eigenstates, as described in Ref.\
\onlinecite{hfetcharge}. In order to illustrate these effects
analytically, the 2DEG is modeled here as a $ \delta $ function
localized at the AlGaN/GaN interface. These induced electrons may be
accounted for by surface charges such as donor states, by deep-level
traps in the AlGaN and GaN layers, or by unintentional dopants.  The
exact origin of the induced 2DEG is still unknown, but its presence is
well-established from previous work.\cite{Ambacher1}  As will be seen
later, representative values for the induced 2DEG are used.  The
1-D Poisson equation then becomes
\begin{eqnarray}
  \label{eq:pois1d}
  \frac{\partial}{\partial z} \left [ \left ( \epsilon +
  \frac{e_{33}^2}{C_{33}} \right ) \frac{\partial \phi}{\partial
  z} \right ] = \frac{\partial P^\textrm{s}}{\partial z} + \nonumber \\ 2
  \frac{\partial}{\partial z} \left [ \left ( e_{31} - e_{33}
  \frac{C_{13}}{C_{33}}
  \right ) \gamma_{xx} \right ] + e_o n^{(2)} \delta (z - t_a) ,
\end{eqnarray}
where $ n^{(2)} $ is the areal 2DEG concentration and $e_o$ is the
magnitude of the electronic charge. Equation (\ref{eq:pois1d}) is
integrated to obtain $ \phi $ subject to the boundary conditions $
\phi$$=$$0 $ at $ z$$=$$0 $ and $ z$$=$$t_a$$+$$t_b$ and also to the
continuity of $ \phi $ and the electric displacement across the
interface.  The condition $\phi = 0$ at the two surfaces presumes that
the polarization charge of the materials is terminated by external
charges and that the applied bias voltage is zero.  The general
solution in each layer, obtained by integrating Eq.\ (\ref{eq:pois1d})
twice, is given by
\begin{equation}
  \label{eq:general}
  \phi =  \frac{P^\textrm{s}}{\epsilon^\prime} z +
  \frac{2 (e_{31} C_{33} - e_{33} C_{13}) \gamma_{xx}}{\epsilon^\prime
  C_{33}} z + \frac{A}{\epsilon^\prime} z + B ,
\end{equation}
where
\begin{equation}
  \label{eq:epsilprime}
  \epsilon^\prime = \epsilon + \frac{e_{33}^2}{C_{33}} ,
\end{equation}
and is layer-dependent because of the varying elastic and
piezoelectric coefficients, and $A$ and $B$ are unknown constants.
Thus there are four unknowns, two in each layer.  The $B$'s are
eliminated by enforcing the boundary conditions $ \phi$$=$$0 $ at $
z$$=$$0 $ and $ z$$=$$t_a$$+$$t_b$.  It should also be noted that $
\gamma_{xx} = 0 $ in the GaN buffer because of the simplifying
assumptions made previously.  The relationship between the two $A$'s
is established from the continuity of the electric displacement and is
found by integrating Eq.\ (\ref{eq:pois1d}) across the AlGaN/GaN
interface (see Fig.\ \ref{fig:schematic}) encompassing the 2DEG.  This
boundary condition gives the relation between the electric fields in
the two layers as
\begin{equation}
  \label{eq:Dcontinue}
  \left . \epsilon^\prime \frac{\partial \phi}{\partial z}
  \right |^{t_a^+}_{t_a^-} = \left . P^\textrm{s} \right
  |^{t_a^+}_{t_a^-} + \left . \frac{2 ( e_{31} C_{33} - e_{33} C_{13})
  \gamma_{xx}}{C_{33}} \right |^{t_a^+}_{t_a^-} + e_o n^{(2)} .
\end{equation}
From Eq.\ (\ref{eq:Dcontinue}), we obtain the relation between the
$A$'s in the respective layers as
\begin{equation}
  \label{eq:A1A2}
  A^\textrm{GaN} = A^\textrm{AlGaN} + e_o n^{(2)} .
\end{equation}
From Eq.\ (\ref{eq:A1A2}) and from the continuity of $\phi$ across the
interface, we can solve for all of the remaining unknowns.

The results are greatly simplified in the limit $
t_b$$\gg$$t_a $, a condition that is met for most HFETs.  In this
limit, the electrostatic potential in the AlGaN layer is given by
\begin{eqnarray}
  \label{eq:phiAlGaN}
  \phi_\textrm{AlGaN} = \frac{P^\textrm{s}_\textrm{AlGaN} -
  P^\textrm{s}_\textrm{GaN} }{\epsilon^\prime} z -
  \frac{e_o n^{(2)}}{\epsilon^\prime} z  + \nonumber \\
  2 \left ( \frac{e_{31} C_{33} - e_{33} C_{13}}{\epsilon^\prime
  C_{33}} \right ) \gamma_{xx} z ,
\end{eqnarray}
and $ \phi $ is constant in the GaN layer.  From Eq.\
(\ref{eq:phiAlGaN}), the electric field in the AlGaN layer is given by
\begin{eqnarray}
  \label{eq:EAlGaN}
  E_z = \frac{P^\textrm{s}_\textrm{GaN} -
  P^\textrm{s}_\textrm{AlGaN} }{\epsilon^\prime} +
  \frac{e_o n^{(2)}}{\epsilon^\prime}   - \nonumber \\
  2 \left ( \frac{e_{31} C_{33} - e_{33} C_{13}}{\epsilon^\prime
  C_{33}} \right ) \gamma_{xx}  .
\end{eqnarray}
The electric field within the standard model is obtained by
replacing $ \epsilon^\prime $ with $ \epsilon $ in Eq.\
(\ref{eq:EAlGaN}).  It is seen that for growth along $[0001]$, the
electric field in the AlGaN from the coupled model is smaller that
of the standard model, since $ e_{33}^2 / C_{33} $ is always
positive. The vertical strain component within the fully-coupled
model is obtained from
\begin{eqnarray}
  \label{eq:ezzcoupled}
  \gamma_{zz} = - 2 \frac{C_{13}}{C_{33}} \gamma_{xx} + \left (
  \frac{2 e_{33} (e_{33}
  C_{13} - e_{31} C_{33})}{C_{33}(\epsilon C_{33} + e^2_{33}) } \right
  ) \gamma_{xx} + \nonumber \\
  \frac{e_{33} (P^\textrm{s}_\textrm{GaN} -
  P^\textrm{s}_\textrm{AlGaN} + e_o n^{(2)})}{\epsilon C_{33} +
  e^2_{33}} .
\end{eqnarray}
The first term  in Eq.\ (\ref{eq:ezzcoupled}) is the typical
result obtained from uncoupled elastic theory.  The remaining
terms are due to electromechanical coupling.

Besides the simple analytical approach described in this work, a
previously-published numerical Schr\"odinger-Poisson
model\cite{hfetcharge,hfetpara} has been modified to include full
coupling.  The elements of the strain tensor enter the
Schr\"odinger equation via deformation potential theory.  One of
these elements, $ \gamma_{zz} $, is now a function of the
electrostatic potential calculated from the Poisson equation.  In
turn, the electrostatic potential is a function of the
piezoelectric polarization, itself a function of $ \gamma_{zz} $.
With each Schr\"odinger-Poisson iteration, the strain must be
updated via Eq.\ (\ref{eq:ezz}).  Also included is the
charge-balance equation as described in Ref.\
\onlinecite{hfetcharge}.  The calculation is regarded as having
converged when the maximum change in the electrostatic potential
between the current and the previous iteration is less than 0.01
meV.

\section{Results and discussion}
\label{sec:results} This section contains calculations used to
compare the uncoupled and coupled models for a model HFET
structure. The presence of the 2DEG and its effect on the problem
are investigated. Table \ref{tab:material} shows the material
parameters\cite{Wright,Yim,Maruska,Tsubouchi,Shur1,Bernardini,Chin}
used in the calculations.  The signs of the polarization
parameters are defined in relation to the [0001] direction: a
negative sign means that the vector is in the $ [ 000 \bar{1} ] $
direction.

\begin{figure}
  \begin{center}
    \includegraphics[width=3.2in]{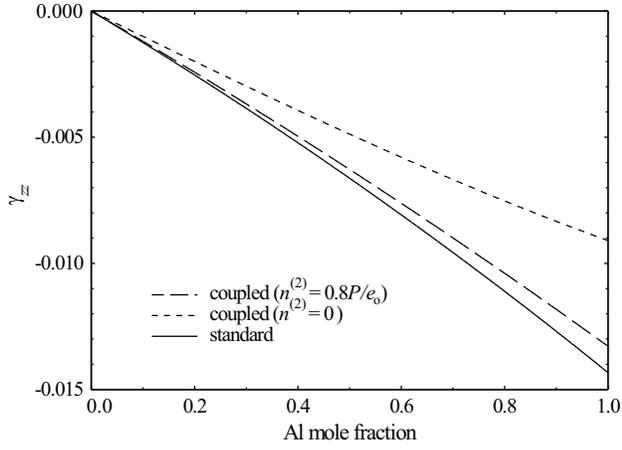}
  \end{center}
  \caption{Calculated out-of-plane strain for the standard and coupled
    models.  Two coupled cases are shown, one without free electrons
    and one with free electrons.}
  \label{fig:poisson}
\end{figure}

\begin{figure}
  \begin{center}
    \includegraphics[width=3.2in]{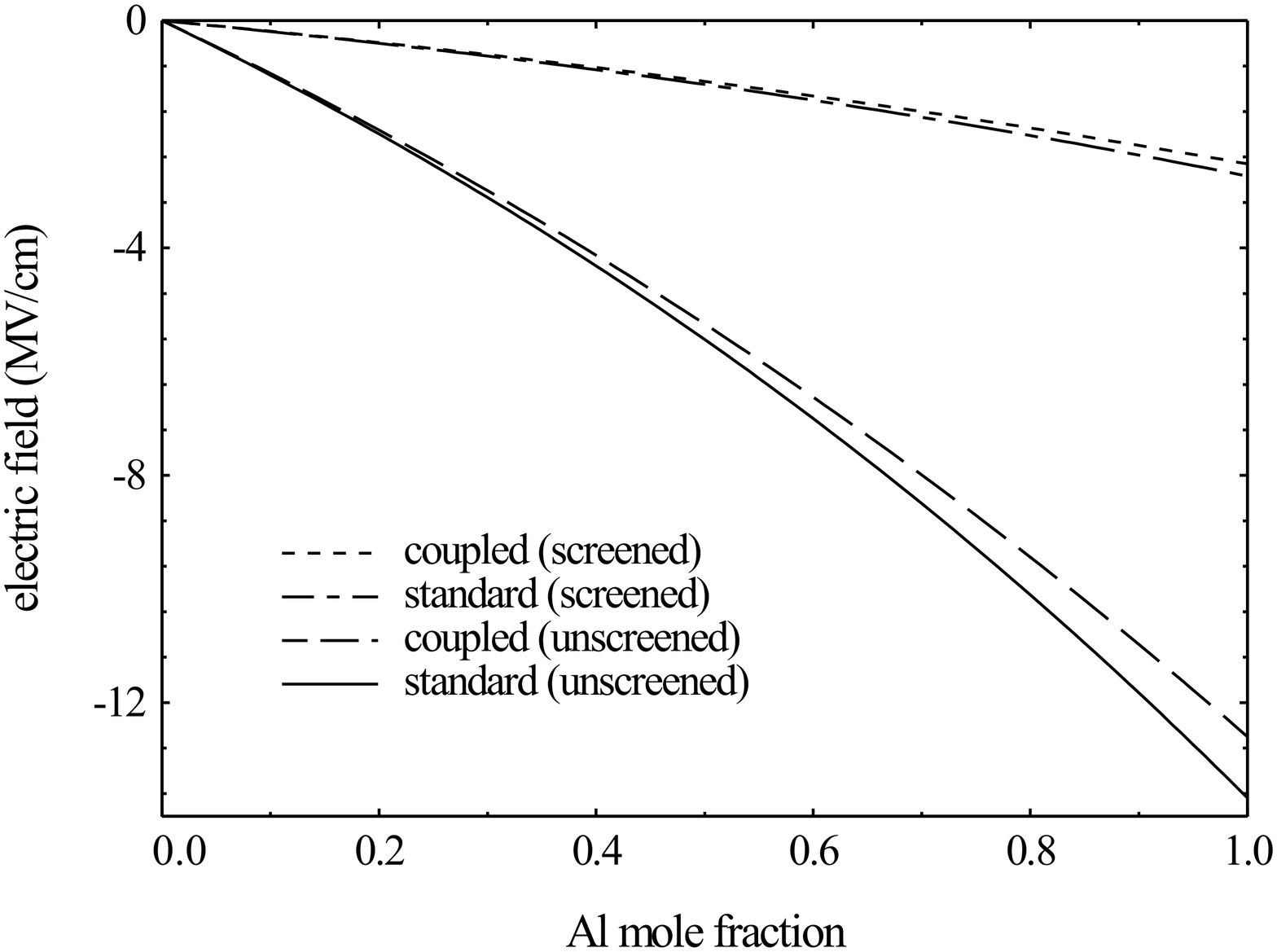}
  \end{center}
  \caption{Electric field in the AlGaN layer for the standard and
    coupled models with and without 2DEG screening.  For the screened
    cases, $ n^{(2)} = 0.8 P $.}
  \label{fig:efield}
\end{figure}

The calculated strain for both the standard and fully-coupled models
as a function of the Al fraction of the barrier layer is shown in
Fig.\ \ref{fig:poisson}.  It is seen that the electromechanical
coupling changes $ \gamma_{zz} $ significantly in the absence of a
2DEG. It is also noted that the coupling reduces $ \gamma_{zz} $
relative to the standard model. This result is evident from Eq.\
(\ref{eq:ezz}) in which it is seen that the coupling opposes the
contraction of material along the $c$ axis when there is tensile
strain in the $c$ plane.  A realistic comparison between the two
approaches should include the mobile charge drawn to the interface by
the polarization space charge. The net polarization space charge at
the interface is given by $
P$$=$$[P^\textrm{s}_{\textrm{GaN}}$$-$$P^\textrm{s}_{\textrm{AlGaN}}$$-$$
P^\textrm{p}_{\textrm{AlGaN}}] $, and for a 30\% Al alloy composition
is approximately $ 1.63 \times 10^{13}\ e_o / \textrm{cm}^2 $. The
2DEG can be expected to neutralize the space charge to some extent.
In Fig.\ \ref{fig:poisson}, $ n^{(2)} $ is taken to be $ 0.8 P/e_o $,
although it will be seen shortly from a Schr\"odinger-Poisson solution
that the neutrality factor is closer to 0.9.  With screening from the
2DEG included, the discrepancy between the two models is reduced. This
result is due to the reduction in the polarization-induced electric
field caused by screening and is shown more clearly in Fig.\
\ref{fig:efield}.  The magnitude of the 2DEG chosen here is
representative of the induced electrons seen in actual HFET
devices.\cite{Ambacher1} We do not, however, explore the origin of the
2DEG in this simple approach, as the only purpose in including the
2DEG is to show that electron screening has a strong influence on the
electromechanical coupling.

\begin{figure}
  \begin{center}
    \includegraphics[width=3.2in]{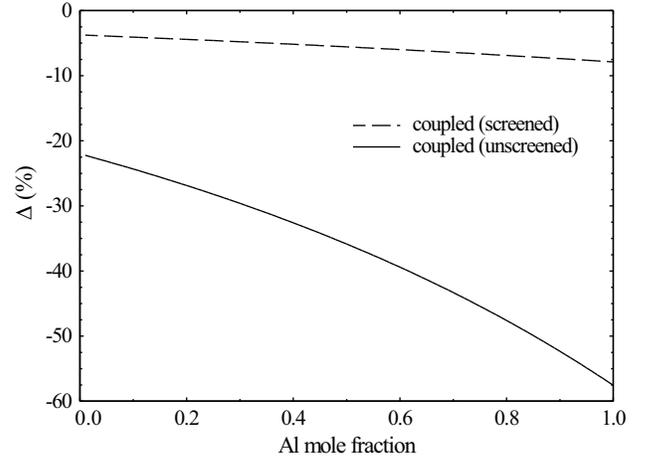}
  \end{center}
  \caption{Deviation of the out-of-plane strain for two coupled cases, with
    and without free-carrier screening.}
  \label{fig:poiserror}
\end{figure}

Next we examine the deviation $\Delta$ between the coupled and
uncoupled models. The deviation in the out-of-plane strain
component $\gamma_{zz}$ is defined as
\begin{equation}
  \label{eq:error}
  \Delta =  \frac{\gamma_{zz} -
  \gamma_{zz}^\textrm{uncoupled}}{\gamma_{zz}}
\end{equation}
where $ \gamma_{zz} $ is given by Eq.\ (\ref{eq:ezzcoupled}) for
both the screened and unscreened cases. The calculated $\Delta$ is
shown in Fig. \ref{fig:poiserror}. For a typical HFET with an Al
mole fraction of 0.3, the error is about 30\% if the effect of the
2DEG is neglected, as compared to 17\% for free-standing plates of
AlN. It is clear that if the AlGaN layer is considered in
isolation, the standard model gives a large error in the strain
tensor.  In an HFET structure, however, the importance of the
AlGaN/GaN interface cannot be overestimated. This interface
introduces additional physics, the most important being the 2DEG.
Effectively, the presence of a large 2DEG restores the validity of
the standard model.

\begin{figure}
  \begin{center}
    \includegraphics[width=3.2in]{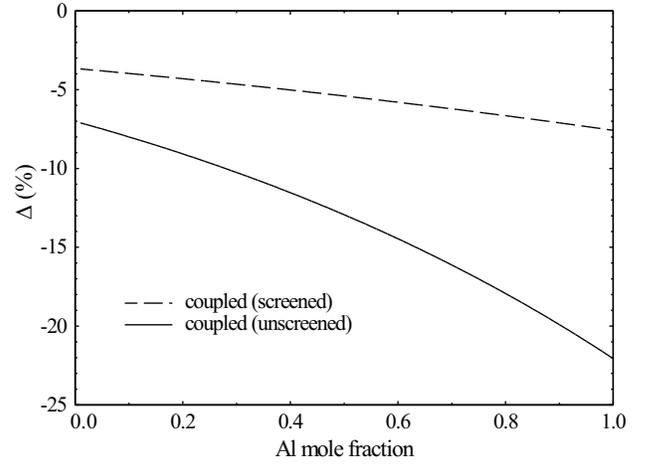}
  \end{center}
  \caption{Deviation of the out-of-plane strain for two coupled cases, with
    and without free-carrier screening with the spontaneous
    polarizaion set to zero.}
  \label{fig:poiserrornops}
\end{figure}

Fig.\ \ref{fig:poiserrornops} shows the difference in
$\gamma_{zz}$ when the spontaneous polarization is excluded.  It
is well-known that the spontaneous polarization is the larger of
the two polarizations in the AlGaN/GaN material system.  For
example, for a Al fraction of 0.3, the space charge induced by the
discontinuity of the spontaneous polarization is about $ 9.8
\times 10^{12}\ e_o / \textrm{cm}^{2} $ as compared to about $ 6.7
\times 10^{12}\ e_o / \textrm{cm}^{2} $ for the piezoelectric
polarization. The difference increases for higher Al fractions.
Accordingly, with just the piezoelectric polarization included,
the electromechanical coupling is underestimated.  This is because
the electric field is reduced compared to Fig.\ \ref{fig:efield}.
Once again, however, the screening from the 2DEG reduces the
coupling.

\begin{figure}
  \begin{center}
    \includegraphics[width=3.2in]{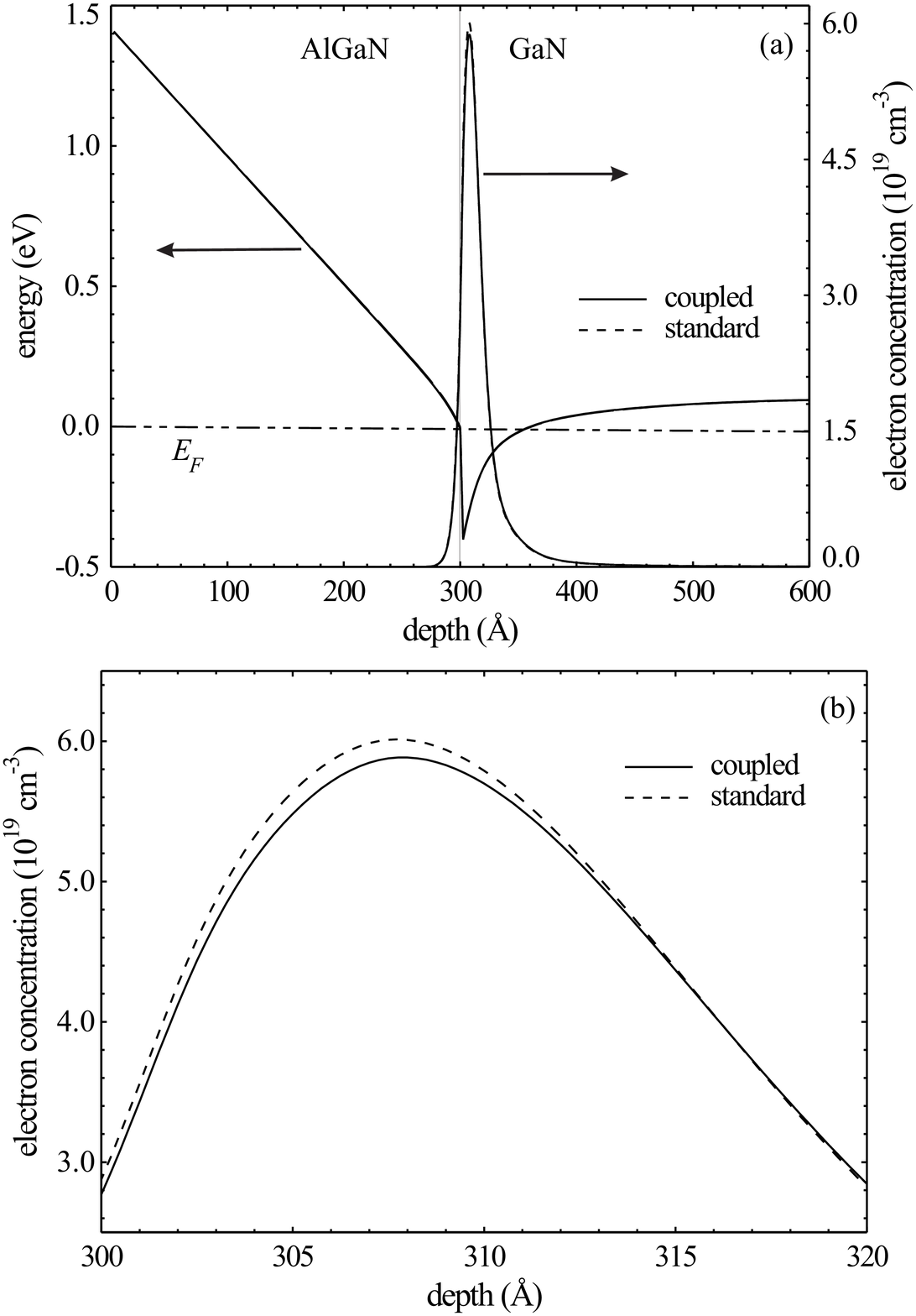}
  \end{center}
  \caption{(a) Calculated conduction band edge (left $y$-axis) and free
    electron distribution (right $y$-axis) along a $c$-axis slice of a
    representative HFET structure at 300 K.  The $ \textrm{Al}_{0.3}
    \textrm{Ga}_{0.7} $N cap layer is 300 \AA{} thick.  The dot-dashed
    line is the Fermi energy.  The surface of the structure is at the
    left extremity and the interior towards the right.  The the 2DEG
    is assumed to originate from surface states.  The standard (solid
    line) and coupled (dashed line) electron distributions are shown.
    (b) The free electron distribution in (a) plotted over a narrower
    region of the structure to show the difference between the coupled
    and standard cases more clearly.}
  \label{fig:self}
\end{figure}

Fig.\ \ref{fig:self}(a) shows the conduction band edge and the
free electron distribution obtained from the Schr\"odinger-Poisson
calculation that includes the full electromechanical coupling. The
structure is a standard HFET design consisting of 300 \AA{} of $
\textrm{Al}_{0.3} \textrm{Ga}_{0.7} $N on a thick GaN buffer. The
2DEG is assumed to come from surface donor states\cite{hfetcharge}
having an activation energy of 1.4 eV below the conduction band
edge at the surface, effectively pinning the surface potential at
1.4 V. The resulting 2DEG concentration after convergence is about
$ 1.47 \times 10^{13} $ $ \textrm{cm}^{-2} $. This magnitude is
about $ 0.9 P/e_o $, almost neutralizing the polarization-induced
space charge.  As an example, the electric field in the $
\textrm{Al}_{0.3} \textrm{Ga}_{0.7} $N layer in the absence of a
2DEG is about 3.1 MV/cm from Fig.\ \ref{fig:efield}. The
fully-coupled Schr\"odinger-Poisson model gives a field of about
0.45 MV/cm and the standard Schr\"odinger-Poisson model a field of
about 0.46 MV/cm.  Within numerical errors, the 2DEG is almost
unchanged between the coupled and standard models as shown more
clearly in Fig.\ \ref{fig:self}(b). Consequently, there is little
change in the conduction band edge between the two models. The
calculated eigenstates from the Schr\"odinger equation and the
electrostatic potential from the Poisson equation show little
change relative to the standard model after both models have
converged to a solution.

\section{Summary and conclusions}
\label{sec:summary} In summary, a fully-coupled electromechanical
model has been presented for the strain in AlGaN/GaN HFETs.  The
model goes beyond the standard Hooke's law description and uses
the fully-coupled equation of state for the stress-strain
relationship.  Using a simple analytical model for the HFET, it is
shown that the present model gives rise to significant changes in
the strain field and, consequently, to the electronic properties
of the device in relation to the standard model in the absence of
free charges.  When free charges are present in the form of a
2DEG, the built-in electric field is reduced through the screening
of the fixed space charge.  This effect, in turn, reduces the
electromechanical coupling and brings the results from the
standard and fully-coupled models into closer agreement.

\acknowledgments The work of BJ was partially supported by the Air
Force Office of Scientific Research (AFOSR) and performed at Air Force
Research Laboratory, Materials and Manufacturing Directorate
(AFRL/MLP), Wright Patterson Air Force Base under USAF Contract No.\
F33615-00-C-5402.

\bibliography{electromech}
\newpage
\begin{table}
\caption{Strain-related material parameters used in the present model.
  The elastic stiffness constants are in units of GPa and the
  piezoelectric stress constants and spontaneous polarization in units
  of C/$\textrm{m}^2$.}
  \label{tab:material}
\begin{tabular}{lcccccccc}
Material & $C_{13}$ & $C_{33}$ & $a$ (\AA{}) & $ e_{31} $ & $
e_{33} $ & $ P_\textrm{spont} $ & $ \epsilon / \epsilon_0 $ \\
\hline AlN & 108\tablenotemark[1] & 373\tablenotemark[1] &
3.112\tablenotemark[2] & $-$0.58\tablenotemark[4] &
1.55\tablenotemark[4] & $-$0.081\tablenotemark[6] &
8.5\tablenotemark[7] \\
GaN & 103\tablenotemark[1] & 405\tablenotemark[1] &
3.189\tablenotemark[3] & $-$0.36\tablenotemark[5] &
1\tablenotemark[5] & $-$0.029\tablenotemark[6] & 10\tablenotemark[7] \\
\end{tabular}
\tablenotetext[1]{Reference \onlinecite{Wright}.}
\tablenotetext[2]{Reference \onlinecite{Yim}.}
\tablenotetext[3]{Reference \onlinecite{Maruska}.}
\tablenotetext[4]{Reference \onlinecite{Tsubouchi}.}
\tablenotetext[5]{Reference \onlinecite{Shur1}.}
\tablenotetext[6]{Reference \onlinecite{Bernardini}.}
\tablenotetext[7]{Reference \onlinecite{Chin}.}
\tablenotetext[8]{Contracted index notation:
$C_{13}$$=$$C_{xxzz}$$=$$C_{yyzz}$$=$$C_{zzxx}$$=$$C_{zzyy}$,
$C_{33}$$=$$C_{zzzz}$, $e_{31}$$=$$e_{zxx}$$=$$e_{zyy}$, and
$e_{33}$$=$$e_{zzz}$.}
\end{table}

\end{document}